# A multi-purpose reciprocating probe drive system for studying the effect of gas-puffs on edge plasma dynamics in the ADITYA-U tokamak

Kaushlender Singh[1,2], Bharat Hegde[1,2], Ashok K. Kumawat[1,2], Ankit Kumar[1,2], M.S. Khan[1,2], Suman Dolui[1,2], Injamul Hoque[1,2], Tanmay Macwan[3], Sharvil Patel[1,4], Abha Kanik[1,5], Komal Yadav[1,2], Soumitra Banerjee[1,2], Harshita Raj[1,2], Devilal Kumawat[1], Pramila Gautam[1], Rohit Kumar[1], Suman Aich[1,2], Laxmikanta Pradhan[1], Ankit Patel[1], Kalpesh Galodiya[1], Abhijeet Kumar[1], Shwetang Pandya[1], K. M. Patel[1], K. A. Jadeja[1], D.C. Raval[1], R. Tanna[1] and Joydeep Ghosh[1,2]

[1]Institute for Plasma Research, Bhat, Gandhinagar, Gujarat, 382428, India
[2]Homi Bhabha National Institute, Training School Complex, Anushakti Nagar, Mumbai, 400094, India
[3]University of California Los Angeles, Department of Physics and Astronomy Los Angeles, CA, 90095, USA
[4]Pandit Deendayal Energy University, Raisan, Gandhinagar, Gujarat, 382007, India
[5]University of Petroleum and Energy Studies, Bidholi, Uttarakhand, 248007, India

Email: Kaushlender.singh@ipr.res.in

**Abstract:** This article reports the development of a versatile high-speed reciprocating drive system (HRDS) with interchangeable probe heads to characterise the edge plasma region of ADITYA-U ($R_0$ = 75 cm, a = 25 cm) tokamak. This reciprocating probe drive system consisting of Langmuir and magnetic probe-heads, is designed, fabricated, installed, and operated for studying the extent of fuel/impurity gas propagation and its influence on plasma dynamics in the far-edge region inside the last closed magnetic flux surface (LCFS). The HRDS is driven by a highly accurate, easy-to-control, dynamic, brushless, permanently excited synchronous servo motor operated by a PXI-commanded controller. The system is remotely operated and allows for precise control of the speed, acceleration, and distance-travelled of the probe-head on a shot-to-shot basis, facilitating seamless control of operations according to experimental requirements. Using this system, consisting of a linear array of Langmuir probes, measurements of plasma density ($n_e$), temperature ($T_e$), potential ($V_p$), and their fluctuations revealed that the fuel gas-puff impact these mean and fluctuating parameters up to ~ 3 – 4 cm inside the LCFS. Attaching an array of magnetic probes to this system led to measurements of magnetic fluctuations inside the LCFS. The HRDS system is fully operational and serves as an important diagnostic of ADITYA-U tokamak.

## 1 Introduction:

Fuel gases, such as Hydrogen ($H_2$), and/or Deuterium ($D_2$) are injected during the progression of a tokamak discharge to control and maintain the density of the plasma [1]. The simplest way to inject these gases is through fast-opening valves, such as piezoelectric or solenoid valves, operated in pulsed or continuous mode, releasing gas at the plasma boundary. Techniques like Supersonic Molecular Beam Injection (SMBI) [2] and pellet injection (PI) [3] are also available for direct fuelling of the plasma interior. The pulsed gas-injection through the plasma boundary, commonly called 'gas-puffs' are primarily used to enhance plasma density [1], however, they also affect the plasma properties in the edge and scrape-off layer (SOL) region of the plasma. Observations suggest that a gas-puff pulse suppresses the turbulence and the associated particle flux [4]. It decreases the heat-load falling on the plasma facing components (PFCs) causing detached plasma states, which is very essential for a safe operation of a fusion reactor [5]. Furthermore, the gas-puffs give rise to several interesting phenomena such as the core temperature enhancement through cold-pulse propagation [6], control of drift tearing mode rotation frequency [7], and sawtooth stabilization [8] etc. To unveil the effect of gas-puff on the edge plasma, measurements of plasma parameters, such as density, temperature, potential, magnetic fields, and their fluctuations are required.

Langmuir probe (LP) has been the most fundamental and robust diagnostic tool for characterising the edge plasma region in tokamaks since its invention [9]. The major advantage of LP is that it provides local measurements in the edge and the Scrape Off Layer (SOL) and have led to several important discoveries such as presence of 2-D potential structures (blobs) [10–12], plasma detachment [5,13], intermittent nature of heat and particle transport [4,14] along with identification of various instabilities and modes excited in edge plasma region [15–19]. However, as the tokamaks progressed, the enhanced temperature, density in the edge region and the enhanced discharge-duration limited the region of plasma accessed by the LPs, which are pre-fixed in the machine before the initiation of a plasma discharge. The probe-heads of pre-fixed LPs melts down as they are exposed to relatively high-temperature, high-density plasma for relatively longer period and contaminate the main plasma [20]. Furthermore, plasma initiation and current build-up is also hampered by the pre-fixed LPs located inside the limiter radius of a tokamak. Similar is the case for the internal magnetic-probes, which provides the information on plasma current profiles and magnetic fluctuations in the edge-region inside the LCFS. The pre-fixed probe-heads containing the magnetic probes (MPs) also disturbs the overall plasma and gets damaged by prolonged exposure of the plasma.

The above-mentioned limitations of the fixed LPs and internal MPs can be overcome using a reciprocating probe drive system, which enables a controlled exposure, in both time and space, of the probe-head inside the plasma. Thus, keeping both the probe and plasma healthy during the operation. Reciprocating probe drives were designed and operated in several tokamaks. In the TEXT tokamak, temporal and spatial profiles of edge plasma parameters ($n_e, \phi_e, T_e$) and their fluctuations were measured by scanning the edge plasma using a pneumatically driven four pin Langmuir probe [21]. In the TJ-II stellarator, the edge plasma region was scanned using a pneumatically driven Langmuir probe [22]. Similar studies of edge plasma parameters have been carried out in the HL-2A, DIII-D, and NSTX tokamaks, using pneumatic systems for fast LP motion [23–25]. To achieve accurate and flexible control on the position and velocity with high repeatability, a servo motor-driven reciprocating probe system was developed and operated in the EAST, ALCATOR C-Mod and TCV tokamaks [20,26,27].

In addition, a magnetically driven reciprocating probe drive system has been used recently to pop-up Langmuir probes for studying the divertor plasma of the W7-X [28]. Although, several reciprocating probe drive systems were designed and operated in different tokamaks, every system design is unique in terms of requirements of velocity, stroke-length, mass etc. of the probe-head along with the machine port-availability, and vacuum constraints. Also, in the servo-motors based drive systems, the motors either need to be proper shielded for magnetic fields or they need to be kept far away from the machine, demanding bigger bellows with bigger stroke-lengths. Hence, the design, fabrication, and operation of such a system for any new machine remains challenging and noteworthy.

In ADITYA-U tokamak [13] (Figure 1), several experiments have been carried out to study the effect of fuel gas-puffs on both the mean and fluctuating edge plasma parameters [4,6–8] primarily using pre-fixed Langmuir and magnetic probes. It has been observed that the pre-fixed probe-head can only be placed up to ~ 1 cm inside the limiter radius (r = 25 cm), i.e., till r = 24 cm ($\rho$ ~ 0.96), before they got severely damaged and impact the plasma production and overall discharge characteristics. However, for gaining knowledge of radial extent of the impact of gas-puffs inside the LCFS, measurements are required far inside the LCFS. To make it possible, a servo motor-based, high-speed reciprocating drive system (HRDS) with interchangeable probe-head, capable of real-time motion during the plasma discharge, has been designed, fabricated, installed, and operated – a first of its kind in ADITYA-U tokamak.

The HRDS for ADITYA-U is designed by considering a usage of a relatively longer bellow with a ~ 50 cm stoke-length to avoid stray magnetic field affecting the servo-motor operation in a constricted space surrounded by machine support structures, a necessity of a gate-valve for changing probe-heads without breaking the vacuum along with managing the electrical and vacuum isolation from the vacuum vessel. The design further includes the limitation posed by the discharge duration of ~ 200 – 400 ms. The system enables the measurement of spatial and temporal profiles of edge plasma parameters up to $\rho$ ~ 0.8. With a LP probe head consisting of 8 probes, the density ($n_e$), temperature ($T_e$), and potential ($V_p$), as well as their fluctuations are measured in presence and absence of gas-puffs. The results indicated that the gas-puff flattens the edge density and temperature profile in the edge region, reducing the fluctuations in these quantities and lead to substantial changes in overall plasma performance including modifications of core temperature and confinement. With internal magnetic probes, the reduction in magnetic fluctuations is also observed with the gas-injection [7].

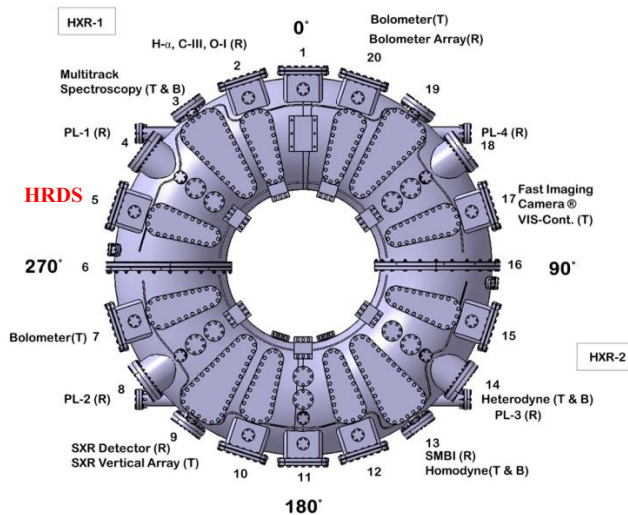

**Table 1: Typical ADITYA-U Parameters**

| Major Radius | 0.75 m |
| --- | --- |
| Minor Radius | 0.25 m |
| $I_p$ (Plasma Current) | 100-200 kA |
| (Toroidal Magnetic field) $B_\phi$ | 0.8-1.5 T |
| $n_e$ (Chord average) | 2-5 x $10^{19}$ m$^{-3}$ |
| $T_{e\_core}$ | 200-500 eV |
| Discharge duration | 100 - 400 ms |
| Edge density ($n_{edge}$) | 1-5 × $10^{18}$ m$^{-3}$ |
| Edge temperature ($T_{edge}$) | 5-20 eV |

*Figure 1: Top-view of the ADITYA-U Tokamak demonstrating various diagnostics locations; Table 1: Shows the typical ADITYA-U parameters.*

Section 2 explains the optimized mechanical design of the HRDS overcoming several technical challenges. The operation and control mechanism of the HRD system is elucidated in section 3 along with its laboratory testing and probe head design. The operation of HRD LP system in glow discharge cleaning (GDC) plasma and in tokamak plasma discharges is presented in Section 4. Section 5 describes the results obtained with HRD LP and MP systems in presence and absence of gas-puffs. The paper is summarized in section 6.

**2. Mechanical Design of HRDS for ADITYA-U:**

Designing a multi-purpose reciprocating drive system for complex machines like tokamaks poses number of challenges, with each machine imposing its own unique set of requirements. Key factors that vary between tokamaks include the availability of space, particularly concerning port location and size, the duration of plasma discharges, the desired spatial and temporal resolution for data collection, and the specific diagnostic parameters to be studied. These considerations are critical for the ADITYA-U tokamak, where the design of the reciprocating drive system must address the following challenges.

1. Plasma Discharge Duration: With discharge durations ranging from 100 to 400 ms, the drive mechanism must enable the probe to scan a range of 4-5 cm within 100-120 ms, requiring an average probe head velocity of 0.4-0.6 m/s.

2. Ultra-High Vacuum Conditions: All components of the system must be compatible with ultra-high vacuum (UHV) conditions ($10^{-8}$ Torr), with a permissible leak rate as low as $10^{-10}$ mbar·l/s to maintain the integrity of the overall vacuum environment of ADITYA-U.

3. Probe Head Flexibility and Stroke Length: The system should allow easy replacement of probe heads based on experimental requirements, i.e., without affecting the overall system vacuum, along with a mechanism for replacing damaged probes. This demands usage of gate-valve and bellow with higher stroke lengths for effective operation.

4. Mechanical Structure and Load Capacity: The structure must be designed to support a total weight of 45 kg, with sufficient strength to handle the momentum of the probe assembly, which reaches approximately 10 kg·m/s.

5. Remote Control and User Interface: Given restricted access to the tokamak hall during operations, remote control of the drive system is essential. A user-friendly interface must allow operators to pre-program parameters such as speed, acceleration, and scan length for the HRDS.

6. Durability and Maintenance of Probe Heads: Probe heads must be capable of withstanding high-temperature plasma environments and feature a modular, easy-to-assemble design for rapid replacement in the event of damage, ensuring minimal disruption to experimental operations.

7. Issues with stray magnetic fields: The driving motor needs to be protected from the stray magnetic field present in the vicinity of the tokamak during operation.

Meeting these requirements is essential to ensure the system operates effectively and experimental explorations of gas puff induced edge plasma dynamics is possible. Considering the above-mentioned contemplations, a HRD system is designed for ADITYA-U tokamak. Figure 2a presents the CATIA drawing of the mechanical design for the high-speed reciprocating drive system (HRDS). Point '1' in Figure 2a (side view of HRDS) represents the ADITYA-U vacuum vessel and the installation port for the HRDS through a 63 CF gate valve (VAT 48.2 XHV Manual DN 63). At this location, the HRDS mechanical assembly and vacuum vessel are electrically isolated using a PTFE (Polytetrafluoroethylene) gasket and bushes, providing an isolation resistance of over 10 MΩ at 100 V DC. This prevents electrical ground loops between the vessel and the HRDS assembly, which is crucial for safety during tokamak operation in the event of a fault and for the probe-signal contamination. A long (1.4 m) stainless steel (SS) edge-welded bellow, with a stroke length of 60 cm, using aluminium gaskets is attached to the other side of the gate-valve using a 63 CF SS 'T' junction in which a vacuum pumping port is provided for evacuating the entire HRD system. A long SS cylindrical shaft (internal diameter = 10 mm and outer diameter =14 mm) is used to hold the probe-head. The shaft is welded to the center-hole (diameter=14 mm) of a 63 CF SS flange, which is mounted on the other end of the bellow at Point 'C' (as shown in Figure 2a in red). As the bellow compresses and expands, the shaft moves in the radial direction placing the probe-head inside and outside the plasma. The shaft-containing flange is terminated to a vacuum electrical feed-through using a 35 CF SS transition flange (at point C). The probe electrical connection runs inside the cylindrical shaft and are brought out of the vacuum using the vacuum electrical feed-through (at point C). The cylindrical shaft is adequately supported along its length inside the mechanical assembly using supports made up of PEEK (Polyether Ether Ketone) material.

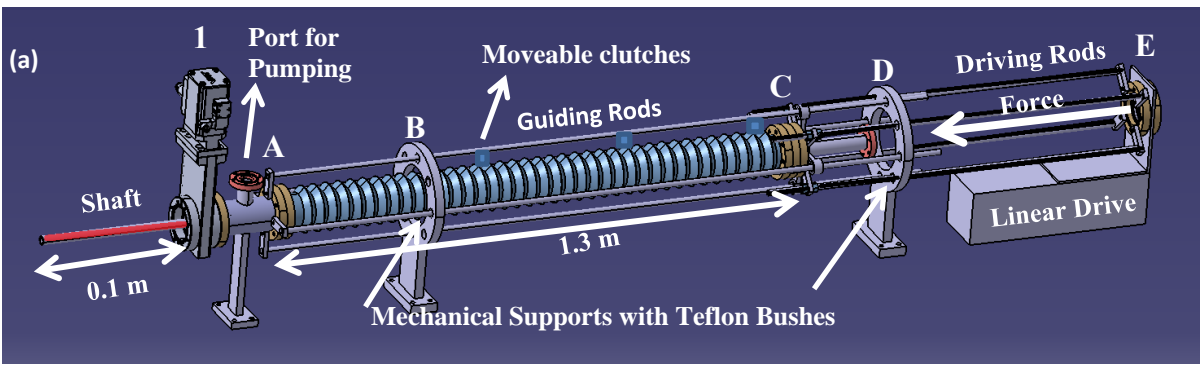

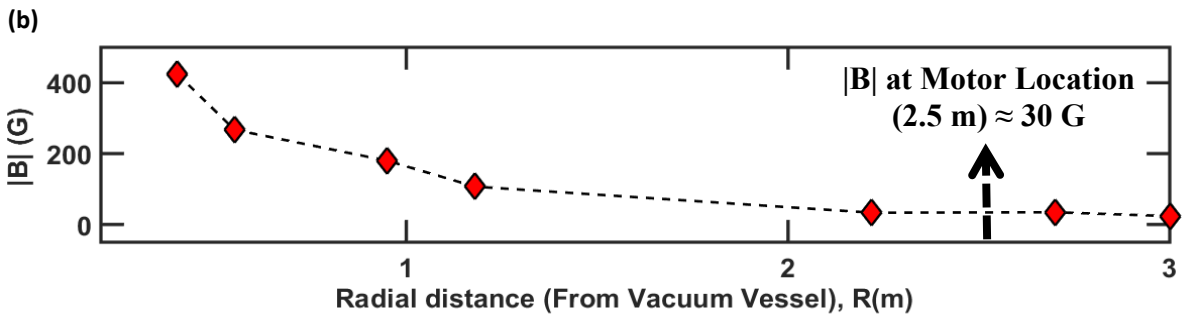

*Figure 2: (a) CATIA drawing of High-speed Reciprocating Drive System for ADITYA-U tokamak (b)Total Magnetic field in the vicinity of ADITYA-U tokamak*

This prevents the shaft from sagging due to gravity and ensures that the probe head moves along the central axis of the vacuum vessel (Z=0). The entire mechanical system is tested in a test chamber for an ultra-low leak rate of $1.1 \times 10^{-10}$ mbar·l/s independent of the motion of the shaft. The other end of the bellow is supported on four 12 mm diameter SS guiding rods (point C in Figure 2a) by four clutches welded on the shaft-containing flange, azimuthally distributed around it. Furthermore, for arresting the sagging of the bellow, it is supported by moveable clutches welded at different points along its length which pass through one of the guide rods. The guide roads are held by two fixed support rings anchored to the base plate at Points 'B' and 'D' and are bolted to the SS "T" at point A. For frictionless movement of the bellow, the junctions of guide-rods and clutches are fitted with industry-grade TEFLON bushes. The motion of the probe-holding shaft is achieved by compressing the edge-welded bellow at Point 'C' using four driving rods (As depicted in Figure 2a in black). One end of these driving rods is bolted on the flange on which the shaft is welded (point C in the figure) and the other is bolted to a linear drive plate (point E), which is fastened to the servo motor. To safeguard the motor operation in ambient magnetic field near the port location necessitates the usage of a long bellow. The strength of ambient magnetic field at various distances from the port location in ADITYA-U is plotted in figure 2b. Considering the field strengths, the servo motor is positioned at ~ 2.5 m away from the vacuum vessel port, where the ambient magnetic field strength during plasma operation is about ~ 30 G. The motor could have been placed near to the port using shielding structures, however, space restriction and the interference of shielding material with tokamak fields did not allow this option.

The specifications of motor for the HRDS are decided based on the static and dynamic torque requirements. The static torque required to hold the assembly in place against the force generated by the differential pressure is estimated first. In static conditions, the primary force arises from the pressure difference between inside ($10^{-8} - 10^{-4}$ Torr) and outside (atmosphere ~ 760 Torr) the bellow. As the system is installed on the vertical midplane port, the system remains parallel to the ground and gravity does not have a direct influence along the direction of the movement. It should be noted that, force due pressure difference ($F_{\nabla P}$) on lateral surface area of the bellow will be perpendicular to the direction of motion and will be counterbalanced by the material strength of the bellow, but, the $F_{\nabla P}$ acting on the cross-sectional area of the bellow will lead to the radial motion, and will have to be balanced by the standstill (holding) torque of the motor. The edge-welded bellow in this case, has an inner diameter of ~ 0.07 m, resulting in an effective cross-sectional area of approximately 0.004 m² (calculated as A=$\pi r^2$). Considering a safety factor of ~ 2, the effective area is taken to be ~ 0.01 m² resulting in a force of ~1000 N that must be counterbalanced by the motor's standstill torque, along with any static friction associated with the mechanical assembly of HRDS. The pitch of the spindle axis (linear actuator) determines the torque of the motor with depends on its effective radius. The spindle axis with a pitch of "P" m, the effective radius, "$r_m$" (the rotational equivalent for torque calculations i.e. $\tau = F \cdot r_m$) is given by $P/2\pi$, resulting in a standstill torque of $\tau_{standstill} = F_{\Delta p} \cdot r_m = F_{\nabla P} \cdot \frac{P}{2\pi}$ ~ 6 Nm for P = 20 mm.

The required torque to move the assembly at the desired speed (~ 0.5 m/s) to scan a distance of ~ 5 cm in 100 ms, requires an acceleration of 10 m/s² for a mass of 10 kg resulting in a driving force ~ 100 N which corresponding to a torque of ~1 Nm. Although friction is reduced using PTFE linear motion bearings, factors such as the recirculating ball bearing systems in the linear actuator, contact between the HRDS assembly and guiding rods, the driving rods with support structure "D", and the interaction between the moving rod and the PEEK support will impact the torque calculations. Friction will reduce the standstill torque but increase the torque needed for motion. Based on these calculations, a servo motor paired with a spindle axis with a recirculating ball bearing guide meets the requirements. A FESTO make dynamic, brushless, permanently excited synchronous servo motor (model: EMMS-AS-100-L-HS-RSB) which can be operated in ambient temperature ranges of -100 to 400 C with a relative humidity of 0 to 90% with a standstill torque of 10.94 Nm, nominal Torque of 7.51 Nm, and peak torque of 39.8 Nm is used to drive the reciprocating system. Optimal operation conditions for the minimum time required for probe motion has been achieved at an average velocity and acceleration of 0.4 m/s and 12.5 m/s$^2$ respectively.

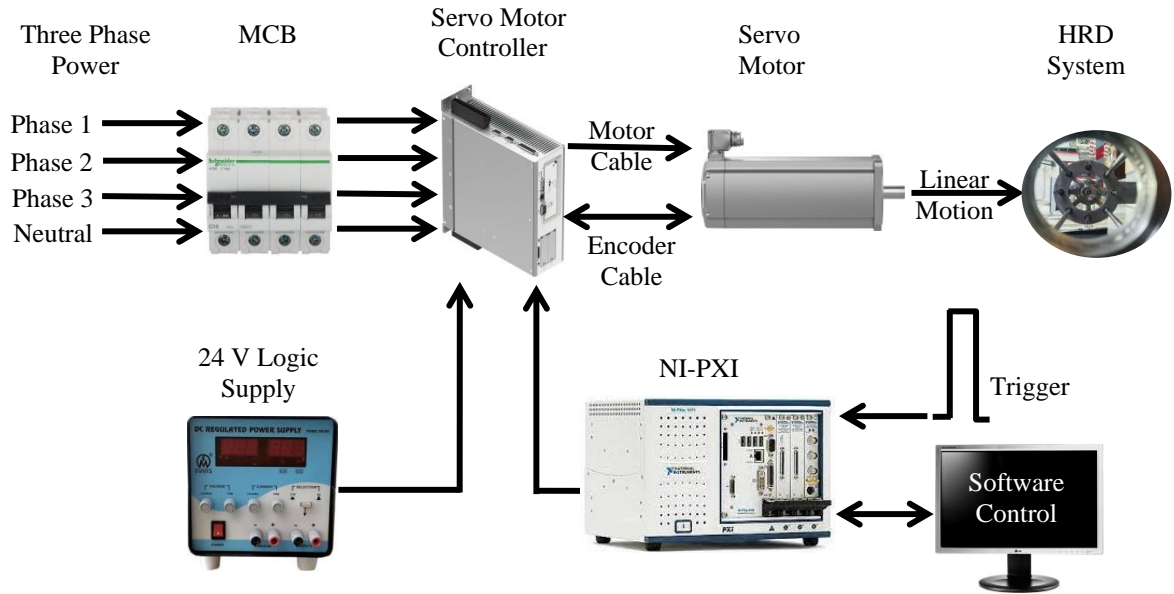

*Figure 3: Schematic for HRD System Operation*

FESTO make controller CMMP-AS-C5-11A-P3-M3 is used for operating the motor, this controller essentially requires main power switch, circuit breaker for protection, and power supply (24 V) for control section, holding break and I/O, comprehensive pinout diagram can be found in reference [29]. The controller is connected to a NI-PXI (National Instruments PCI eXtensions for Instrumentation) via an Ethernet cable for parameter configuration, enabling the user to adjust the velocity, acceleration, and distance of the probe's movement through FESTO software. The synchronized operation with tokamak is initiated by LABVIEW software specifically designed for this purpose. The motor is connected to the controller using two cables: a motor cable for current supply and an encoder cable for signal transmission between the motor sensors and the controller. Figure 3 shows the schematic diagram for motor operation. Three-phase power along with the neutral is fed to the servo motor controller through a manual circuit breaker for safety and switching purposes, along with a 24 V logic supply. These connections enable the controller to supply a nominal operating DC voltage of 565 V to the motor with a current of 3.8 A. The motor operating parameters are programmed via FESTO software and loaded to the controller, which supplies the information to the servo motor. Synchronized operation of the HRDS with ADITYA-U plasma discharges is obtained by operating the HRDS with a trigger mechanism. After receiving the trigger, the motor rotates, and through a spindle axis with a recirculating ball bearing guide (EGC-120-200-BS-25P-KF-0H-MR-GK) with a 20 cm stroke length, 20 mm spindle pitch, maximum speed and torque up to 2 m/s, and 144 Nm respectively can be achieved.

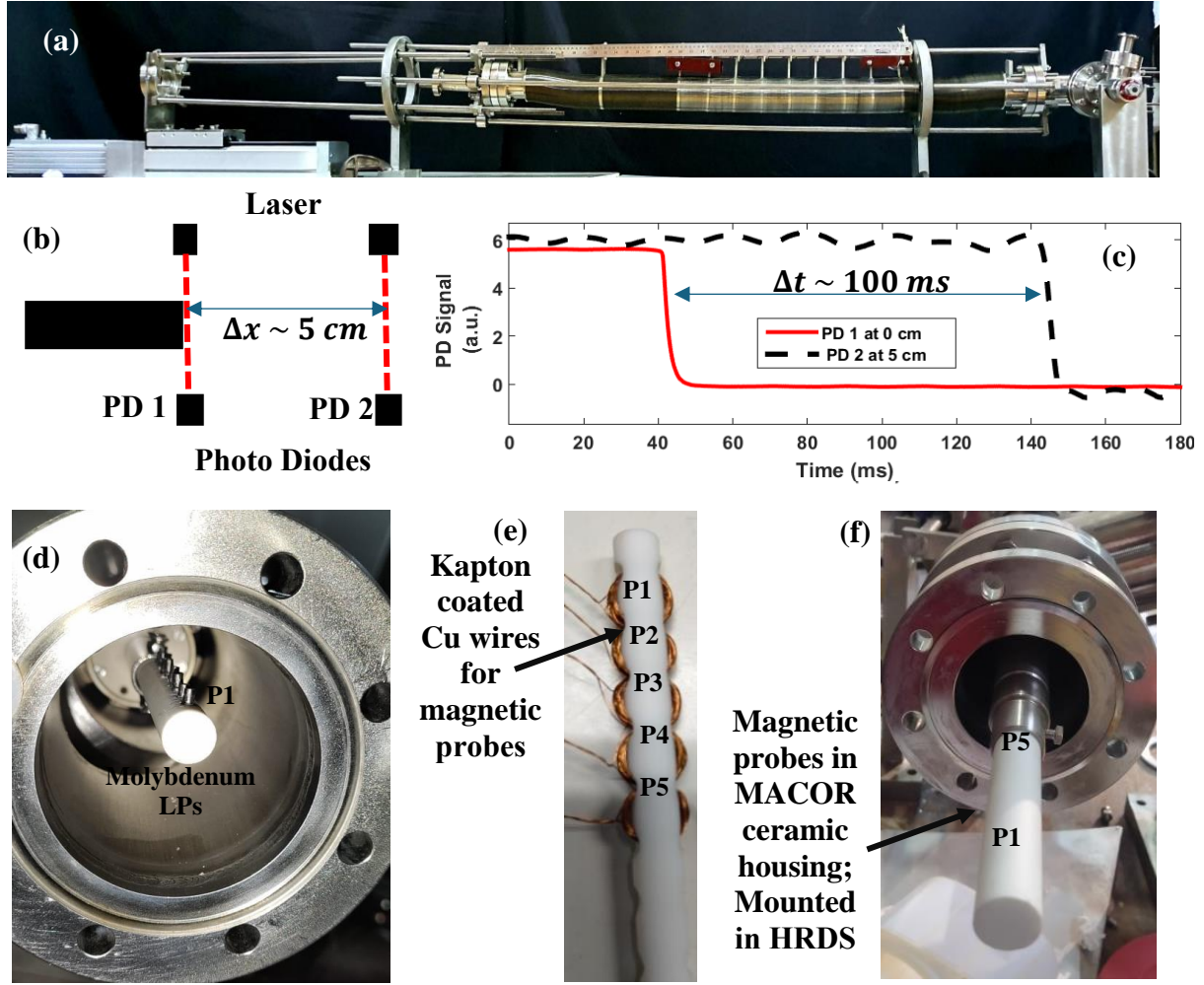

*Figure 4: (a) servo motor-integrated mechanical setup of the HRDS (b) Schematic of laser and photodiode assembly used for speed measurement (c) Signal from photo diode 1 at 0 cm (red solid) and photo diode 2 at 5 cm (black dotted) (d) LP design (e) Magnetic probe (f) Magnetic probes in MACOR ceramic housing; implemented for investigating gas puff induced edge plasma dynamics*

## 3. Operation of HRDS assembly and Probe head designs:

After assembling all the components of HRDS and integrating the servo motor (as shown in Figure 4a), several dedicated test runs have been conducted to ascertain the desired vacuum sealing and high-speed motion of the shaft under both static and dynamic conditions. The average speed of motion has been measured using two pairs of laser and photodiode module positioned 5 cm apart from each other along the linear axis of the moving plate which pushes the bellow. The lasers and photodiodes are mounted perpendicular to the plate thickness (vertical leg of the plate) on either side of its width as shown in figure 4b. The linear movement of the plate interrupts the laser light, causing a dip in its intensity measured by the photodiodes. The average speed of motion of the plate is estimated by dividing the distance between the two photodiodes ($\Delta x$) (Figure 4a) with the time difference ($\Delta t$) (Figure 4b) between the intensity minima measured by the two photodiodes, $v_{avg} = \Delta x/\Delta t$. The desired average speed of motion of ~ 0.5 m/s is achieved at different operating pressures (760 Torr, $10^{-2}$ Torr, and $10^{-8}$ Torr). The velocity remained consistent regardless of pressure variations inside the system. Furthermore, a one-to-one correspondence has been ascertained between the motion of the end-plate pushing the bellow and the motion of the shaft holding the probe-head by measuring the plate and shaft movement separately outside the vacuum. It has been further ensured that the global leak rate of the system (~ $1 \times 10^{-10}$ mbar.l/s) remained unaffected by multiple high-speed reciprocating actions of the shaft. These laboratory tests confirmed the successful operation of the HRDS assembly and validated its installation on tokamak with the Langmuir and magnetic probe-heads.

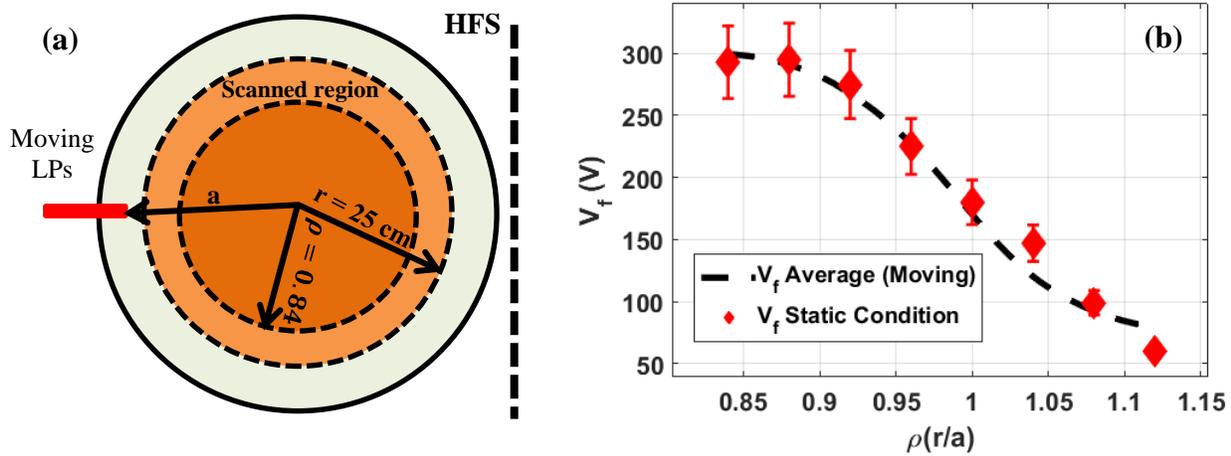

*Figure 5: (a) the schematic of probe motion in GDC plasma, (b) floating potential measured from reciprocating LPs in GDC plasma*

To achieve the primary goal of investigating the effects of gas-puffing on edge plasma turbulence and spatial profiles, the HRDS is installed on ADITYA-U tokamak with an in-house developed rake Langmuir probe (LP). The rake LP contains 8 cylindrical probe-tips, made up of molybdenum, with each probe-tip having a diameter of 2.65 mm (±0.02 mm) and a height of 3.8 mm (±0.2 mm). The probe-tip dimensions are chosen based on their thermal integrity in the long duration plasma discharges (> 200 ms) as they travel to relatively high-temperature zones of the plasma while used in the reciprocating system. The temperature rise of the probe-tips is restricted to ~ 1000°C with an estimated heat load of 5 MW/m². Since, in ADITYA-U tokamak, the radial turbulence decorrelation, density, and temperature scale lengths are greater than 10 mm [4], a spatial distance of 8 mm between each probe is chosen, resulting in a spatial resolution of 8 mm. Figure 4b shows a front view of the HRD system with a rake LP head, consisting of 10 cylindrical probe-tips. The molybdenum probe-tips are brazed with the Kapton coated copper wires, using silver as filler material to keep the resistance of each probe-tip connection < 1 Ω. The electrical connections are brought out of the vacuum using a multi-pin vacuum-electrical feedthrough. High resistance ~ 10-20 MΩ has been ensured between the probe-tips and between the probe-tip and mechanical assembly at 100 V DC. The high insulation resistance significantly reduces the probability of ground loop formation and prevents short circuits among different probe-tips. After collecting data with reciprocating LPs in presence and absence of gas-puff in several discharges (results presented in the next section), a probe-head containing magnetic probes is attached to the HDRS for studying the consequences of gas puffing on magnetic fluctuations. A set of five radially separated magnetic probes has been mounted in the HRDS probe-head. These magnetic probes, which are designed to measure the poloidal magnetic field at 5 different radial locations with a radial resolution of 8 mm. Each probe has a diameter of 8 mm and consists of 50 turns, as depicted in Figure 4c. To ensure protection from plasma exposure, the probes are encased in a MACOR ceramic housing (Figures 4d). The self-inductance ($L_P$) and resistance ($R_P$) of each probe are ~ 15 ± 0.4 µH and ~1.9 ± 0.2 Ω respectively giving a time response $\tau = L_p/R_p \sim 7.5\ \mu s$ corresponding to a frequency response $f = 1/\tau \approx 130$ kHz. The capacitance ($C_P$) of each magnetic probe is ≈ 100 nF, and the resonance frequency of the equivalent LCR circuit $f_{res} = \frac{1}{2\pi\sqrt{L_p C_p}} \approx 129$ kHz.

## 4. Operation of HRD system with plasma:

The HRD system has been installed in the ADITYA-U tokamak with LP-head meeting all the vacuum requirements of the tokamak system. Before operating the HRD system in the tokamak plasma, its operation has been tested in continuous glow-discharge cleaning (GDC) plasma, which is primarily utilized for the conditioning of the vacuum vessel. After pumping out the HRD mechanical assembly, the gate-valve of the HRD system is opened and the first LP is placed inside the vacuum vessel at a desired location outside the limiter radius at r = 28 cm ($\rho = 1.12$). The floating potential of the GDC plasma is measured with the reciprocating the probe up to $\rho = 0.84$. The schematic of the scanned location and probe movement direction in the GDC plasmas is shown in Figure 5a. The probe-tips measured the parameters two times while going-in and coming-out from the GDC plasma. The average of two measurements of floating potential at different radial locations is plotted in figure 5b (Black dotted line). The GDC plasma is produced by placing two electrodes at vacuum vessel centre and biasing it positively with respect to vessel, which is kept at ground potential. Hence, the floating potential increases from ~ 60 to ~300 V while moving inside the GDC plasma towards the centre. While reciprocating, the probe-tip's radial locations are identified from the probe-head velocity measurements using the laser-photodiode pairs. Later, the probe-head is manually placed at the same radial locations and floating potential is measured ( shown by diamond symbols in figure 5b). The reciprocating and manual measurements matches quite well, validating the appropriateness of the probe-head velocity measurement and identification of radial locations of probe-tips during their movement.

Following the successful testing of the HRDS driven LP-head in GDC plasma, it has been operated in plasma discharges of ADITYA-U tokamak and various parameters are measured inside the Last Closed Flux Surface (LCFS). The plasma parameters of two similar discharges, one (#37012) in which the first LP of the HRD system is kept fixed at r = 28 cm and another (#36968) in which the LP is moved inside the LCFS during the discharge, are compared in figure 6. The temporal evolution of loop voltage ($V_L$), plasma current ($I_P$), H-alpha emission intensity ($H_\alpha$), soft X-ray emission intensity (SXR), poloidal magnetic field oscillations ($\tilde{B}_\theta$) of these discharges are shown in figures 6a to 6e, respectively. Figure 6f and 6g presents the temporal evolution of floating potential and plasma density measured using a fixed Langmuir probe kept at r = 26 cm for both the discharges. The pre-filled hydrogen gas is broken down by applying a loop voltage ~ 20 V (figure 6a).

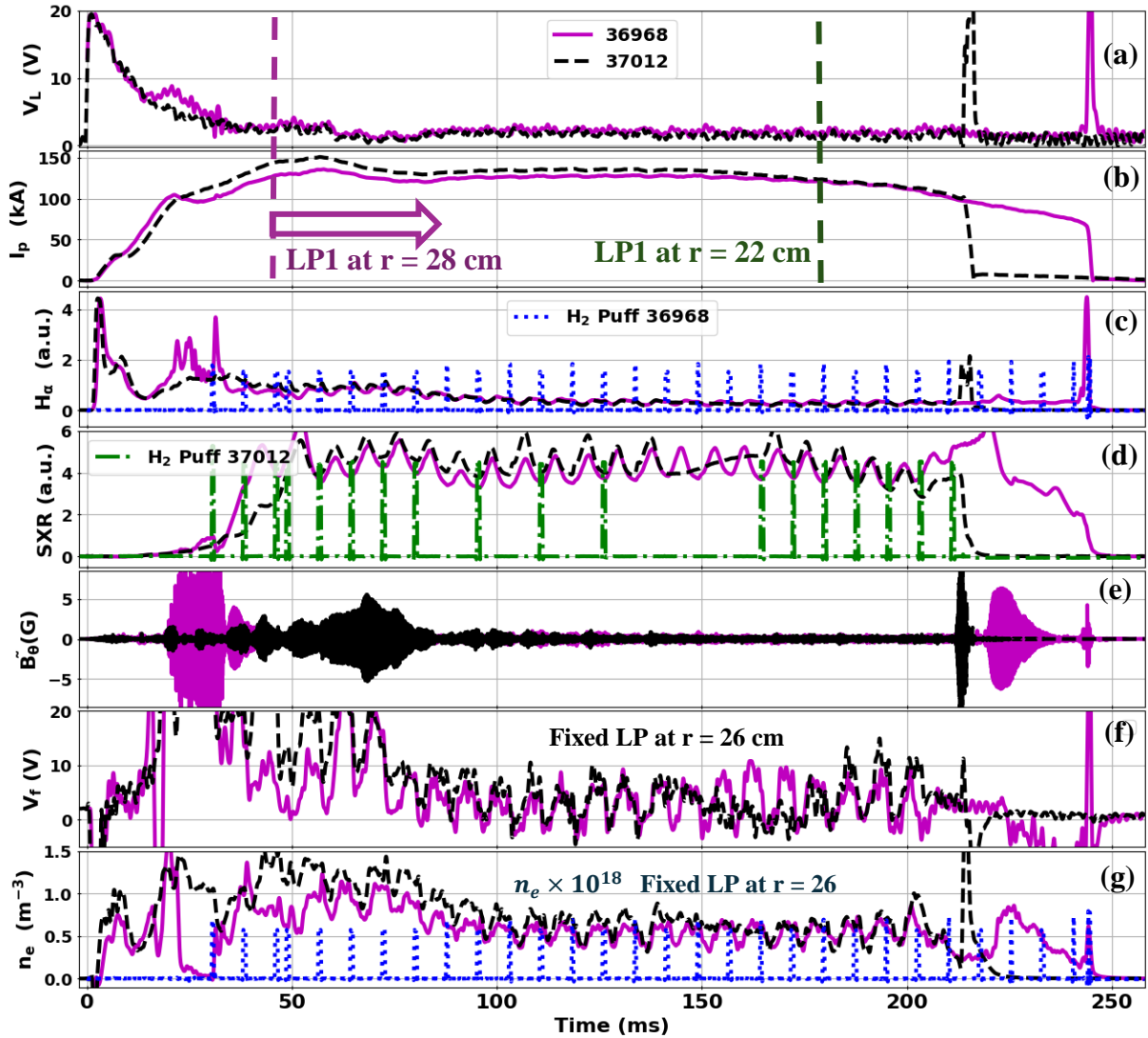

*Figure 6: Comparison between plasma parameters discharge #36968 (magenta solid) and #37012 (Black dotted) (a) loop voltage ($V_L$); (b) plasma current ($I_p$); (c) H-alpha emission intensity ($H_\alpha$);(d) soft X-ray emission intensity (SXR);(e) poloidal magnetic field oscillations ($\tilde{B}_\theta$);(f) floating potential (V), and (g) plasma density ($m^{-3}$) measured from LPs at r = 26 cm*

After the breakdown the plasma current increases and reaches to its flat-top value of ~ 140 kA. At the initiation of both the discharge (t=0), the first LP-tip of HDRS is stationed at r = 28 cm. A trigger, synchronized to the master trigger of the tokamak, initiates the movement of the probe-head at t ~ 40 ms into the flat-top current region of the discharge #36968. No significant variation (< 2%) is observed in the temporal evolutions of plasma current, SXR emission intensity, edge density, and edge plasma potential measured by the fixed LPs remain during the current flat-top phase of the discharge (~ 40 ms to 200 ms) due to the movement of the LP-head. This demonstrates that the LP-head movement using HDRS does not perturb the plasma. Similarly, the magnetic probe head movement also did not perturb the plasma.

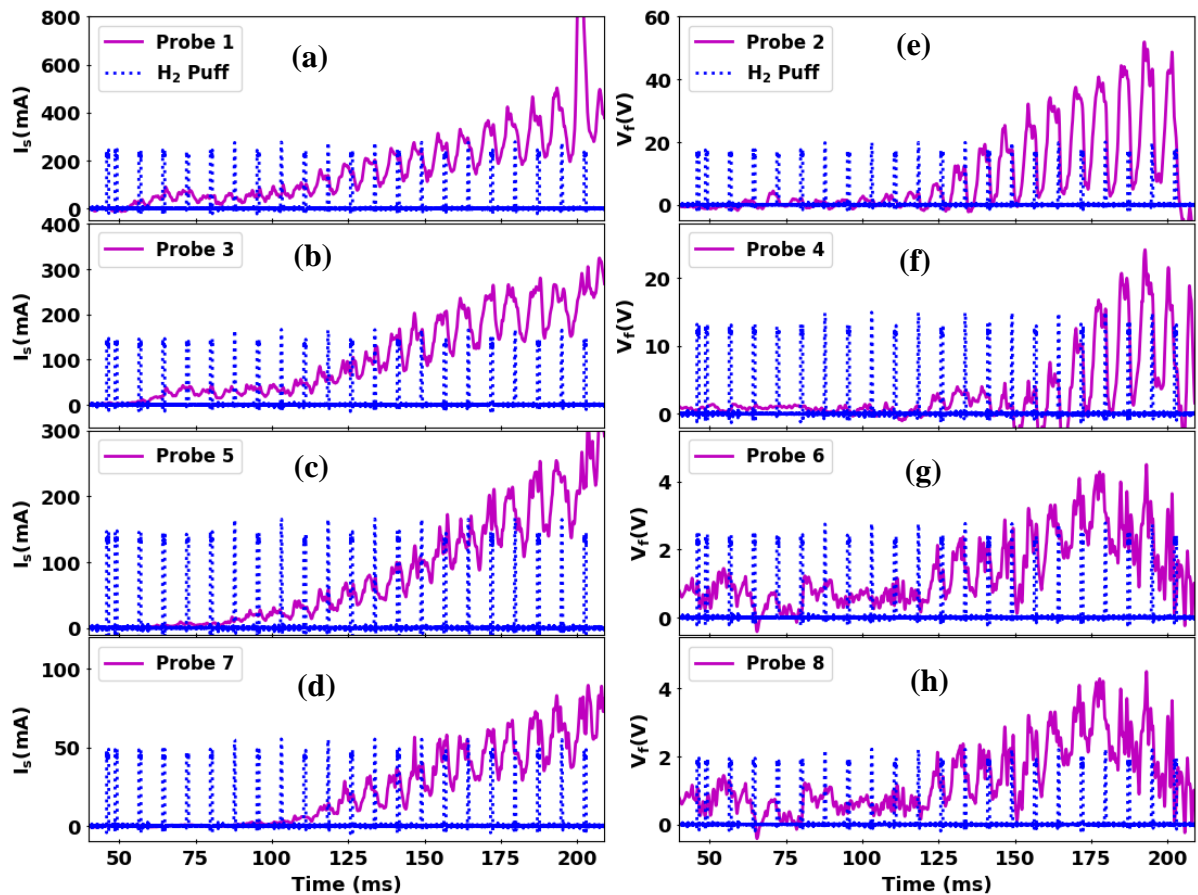

Figures 7: Temporal evolution of ion-saturation current measurement (a) Probe 1; (b) Probe 3; (c) Probe 5; (d) Probe 7; and floating potential measurement (e) Probe 2; (f) Probe 4; (g) Probe 6; (h) Probe 8; during LP motion inside the plasma in the current flat-top phase of the discharge # 36968

## 5. Effect of gas-puffs on edge plasma dynamics:

After thoroughly testing the HRDS system and ensuring the negligible impact of probe-head movement using HRDS on the plasma, the effect of gas-puffing (injection of short bursts of fuel gas) on edge plasma parameters are studied [1,4]. For this experiment, out of the 8 probe-tips of LP-probe head, marked in the ascending order from inside of the vessel to the outside, probe-tips #1, #3, #5, and #7 are used to measure the ion-saturation current, whereas probe-tips #2, #4, #6, and #8 measure the floating potential. Figures 7a – 7h show the temporal evolution of ion-saturation current and floating potential measured by each probe- tip during their motion inside the plasma in the current flat-top phase of the discharge. The time of gas-puff pulses are also shown in the figure 7, by dotted pulses. As the probe-tips move inside the plasma, the mean value of ion saturation current increases as the plasma density increases at inner radial locations away from the LCFS. The mean value of floating potential also increases as the probe-tips move towards the plasma core. Ion-saturation current and floating potential measured with two moving proves (probe-tip #1 and #2) respectively are compared with those measured with fixed probe-tips located at r = 26 cm (one cm outside the limiter radius) in figure 8. Figure 8a shows the comparison of edge plasma density measurement between the moving probe-tip #1 with that of fixed probe-tip.

Before the probe-tip movement of HRDS started, i.e., ~ 40 ms into the discharge, the probe-tip #1 is located at r = 28 cm. Hence the measurement of plasma density is less as compared to fixed probe, which is located at 26 cm. As the probe-tip #1 of HRDS, starts moving, the density increases along its motion towards the core of the plasma. With a probe velocity of ~ 0.5 m/s, it reaches r = 26 cm at ~ 80 ms, and hence it measures the same value of density that is measured by the fixed probe (Figure 8a). Beyond r = 26 cm, the density further increases as the probe-tip moves further in. Similarly, the probe-tip #2 measures the same floating potential that measured by fixed probe when it reaches r = 26 at ~ 90 ms (Figure 8b).

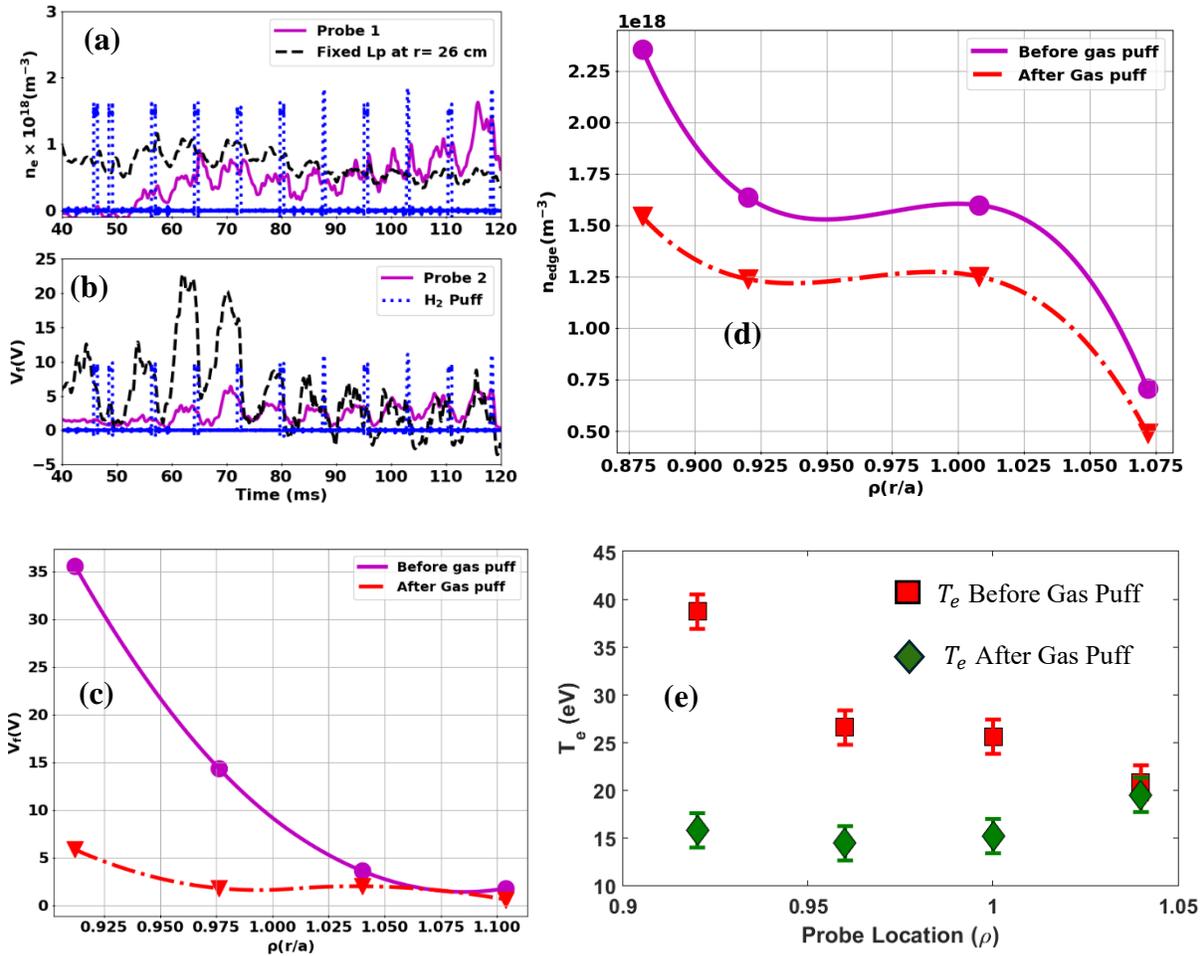

*Figure 8: Comparison of HRD LPs with Fixed LPs measurement (a) for $V_f$ (b) for $n_e$; Effect of fuel puff on spatial plasma profiles measured from HRDS driven LPs (c) for $V_f$; (d) for $n_e$ (plasma discharge #36968); and (e) for $T_e$ (plasma discharge #36971)*

The fixed probe measurements plotted in Figure 6f and 6g shows the effect of gas-puffs on the floating potential and edge plasma density, that with the gas injection, both the density and floating potential decreases sharply immediately after a gas-puff and increases again to regain to pre-gas-puff value before the next gas-puff. All the probe-tips of the HRDS show the similar behaviour (Figure 7). Using the probe-tips of HRDS, the radial profiles of plasma density and floating potential before and after a gas-puff pulse are generated and plotted in figures 8c and 8d respectively, this shows that injected gas affects the plasma parameters up to quite far inside the LCFS, and flattens the radial profiles of both the density and floating potential, influencing the gradient-driven fluctuations in this region of plasma. Similarly flattening of edge temperature profile is also observed due to gas puff injection (Figure 8d).

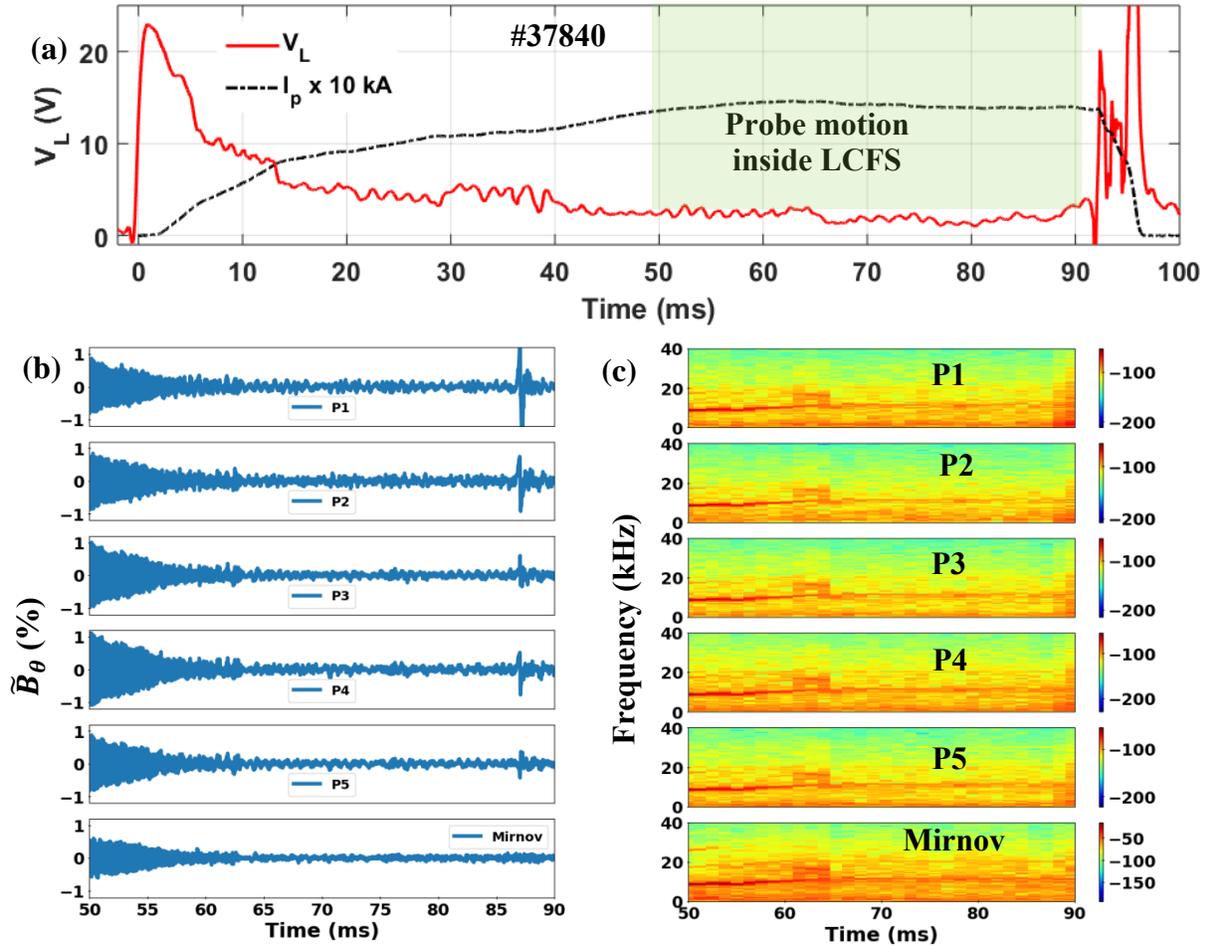

*Figure 9: (a) Plasma current and loop voltage when magnetic probes start moving inside LCFS (b) $\tilde{B}_\theta$ (%) measured from reciprocating magnetic probe and Mirnov probe and their spectrogram for plasma discharge #37840*

After collecting the data with the LP-head of HRDS, it has been replaced with the probe-head containing magnetic probes. The magnetic probes are oriented to measure the poloidal magnetic fields. Figure 9a illustrates the temporal evolution of loop voltage and plasma current for discharge #37840, in which the magnetic-probes are moved using the HRDS. The movement of the probes is initiated at t = 50 ms into the discharge and continued up to t = 90 ms (shaded region in figure 9a). The poloidal magnetic field fluctuations measured with all the magnetic probes present in the HRDS probe-head and by one Mirnov coil (located at r = 28 cm) is shown in figure 9b. In figure 9c, the spectrogram of the poloidal magnetic field fluctuations measured by all the HRDS probes and one Mirnov coil measurement are presented. The time-series and spectrogram of all the HDRS probes compare very well with that of the fixed Mirnov probe.

## 6. Summary and conclusion:

A versatile High-speed Reciprocating Drive System (HRDS) with interchangeable probe-heads has been designed, installed, and operated successfully to study edge plasma dynamics in the ADITYA-U tokamak. To customize the HDRS operation in ADITYA-U, the system is designed by considering several factors specific to ADITYA-U tokamak, such as plasma discharge duration, port dimensions, available peripheral space, ultra-high vacuum requirements, penetration stroke length, static/driving torque requirements, mechanical load-bearing structure etc. Furthermore, attention is being paid to avoid the influence of stray magnetic fields at the system location. The design also facilitates the operation of the system with different probe-heads. Based on the static and moving torque requirements, a brushless, permanently excited synchronous servo motor operated by a PXI-commanded controller is used in the system which provided a precise control of the speed, acceleration. Two laser-photodiode pairs are used to measure the speed of the moving shaft which holds the probe-heads.

After testing the HDRS thoroughly for all its functions in laboratory, it has been installed on ADITYA-U tokamak equipped with a linear array of Langmuir probe-tips. The functionality of the system is further tested by measuring the radial profile of floating potential in glow discharge cleaning plasma. The derived probe-tip locations during the probe-head motion using the laser-photodiode pairs, has been cross-checked by placing the probe-tips at desired location manually in the GDC plasma. Following the successful testing of the HDRS in laboratory and GDC plasma, it has been operated in the tokamak discharge with different probe-tips measuring floating potential, ion-saturation current and the complete I-V characteristics providing radial profiles of radial electric field, density and temperature respectively. Depending upon the plasma duration, the maximum penetration length of the probe-head ~ 5 cm is achieved. The plasma properties did not vary significantly due to the probe-head movement, indicating a minimal perturbation to the plasma. The measurement of floating potential, ion-saturation current and I-V characteristics, in presence and absence of gas-puffs show that a gas-puff of magnitude ~ $10^{18}$ molecules of $H_2$ influences the plasma parameter up to ~ 3 – 4 cm inside the LCFS. The gas-puff leads to a gradual decrease in the mean values of density, temperature, and floating potential (radial electric field) inside the LCFS causing flattening of their radial profiles which affects the magnitude of their fluctuations and subsequently plasma transport in that region [4]. In addition to Langmuir and magnetic probe-heads, the HRDS can be utilized for driving various other probe-heads, such as Mach probes, ball-pen probes, retarding field energy analyser (RFEA) etc. Furthermore, it can also be used for experiments such as electrode biasing, fuel injection inside the LCFS etc

**Acknowledgment and Author Contributions:**

This paper is based on the Ph.D. dissertation research of the first author, who is enrolled at HBNI, Mumbai. The first author is the primary contributor in conceptualizing, designing and development of the HRDS, conducting the experiments, constructing the probes, analysing the data, preparing the original draft and revising the manuscript. Authors are thankful to the IPR workshop and drafting section for their support in fabrication of various parts.

**Author declaration:** Authors have no conflicts to disclose and the data that support the findings of this study are available from the corresponding author upon reasonable request.